\documentclass[runningheads]{svmult}

\usepackage{makeidx}   
\usepackage{graphicx}  
\usepackage{subeqnar}  
\usepackage{multicol}  
\usepackage{physprbb}  
\makeindex             

\begin{document}
\title*{From Intracellular Traffic to a Novel Class\protect\newline of Driven Lattice Gas Models}
\toctitle{From Intracellular Traffic to a Novel Class\protect\newline of Driven Lattice Gas Models}
%
%
\titlerunning{From Intracellular Traffic to a Novel Class of Driven Lattice Gas Models}
%
\author{Hauke Hinsch\inst{1}
\and Roger Kouyos\inst{2}
\and Erwin Frey\inst{1}}
\authorrunning{Hinsch et al.}
%
%
\institute{Arnold Sommerfeld Center and CeNS, Department for Physics,
  Ludwig-Maximilians-Universit\"at M\"unchen, Theresienstrasse 37,
  D-80333 M\"unchen, Germany \and Theoretical Biology,
  Eidgen\"ossische Technische Hochschule Z\"urich,
  Universit\"atsstrasse 16, CH-8092 Z\"urich, Switzerland }

\maketitle              

\begin{abstract}
  Motor proteins are key players in intracellular transport processes
  and biological motion. Theoretical modeling of these systems has
  been achieved by the use of step processes on one-dimensional
  lattices. After a comprehensive introduction to the total asymmetric
  exclusion process and some analytical tools, we will give a review
  on different lines of research attracted to the aspects of this
  systems. We will focus on the generic properties of a coupling
  between the exclusion process and Langmuir bulk kinetics that induce
  topological changes in the phase diagram and multi-phase
  coexistence.
\end{abstract}

\section{Introduction}

The identification of motion as a manifestation of biological life
dates back to the earliest records of science itself. The Greek
physician Erasistratos of Ceos studied biological motion on the length
scale of muscles already in the 3rd century BC. He imagined muscles to
function in the way of a piston contracting and relaxing from
pneumatic origin.  It was not until the invention of the microscope in
the 17th century by van Leeuwenhoek that this theory could be
devalidated with Swammerdams observation that muscles contract at
constant volume.

Concerning biological motion on a microscopic scale, scientists
favored concepts of ``living forces'' for many centuries until this was
finally ruled out by the observations of the Scottish botanist Robert
Brown in 1828 who found all kind of matter to undergo erratic motion
in suspensions.  A satisfactory explanation was provided by Einstein
in 1905 by the interaction with thermally fluctuating molecules in the
surroundings. However, the molecular details remained unknown in the
fog of low microscope resolution. Modern experimental techniques
\cite{metha} have lately revealed the causes of sub-cellular motion
and transport.

Today we know that every use of our muscles is the collective effort
of a class of proteins called myosin that ``walk'' on actin filaments.
Generally spoken, we refer to all proteins that convert the chemical
energy of ATP (adenosine-triphospate) in a hydrolysis reaction into
mechanical work as molecular motors. These motors are highly
specialized in their tasks and occur in a large variety: ribosomes
move along mRNA strands while translating the codons into proteins,
dynein is responsible for cilia motion and axonal transport, and
kinesins play a key role in cytoskeletal traffic and spindle formation
(for an overview see \cite{howard} and references therein).

While the exact details of the molecular structure and function of
motor proteins~\cite{schliwa} remain a topic of ongoing research, on a
different level attention was drawn to phenomena that arise out of the
collective interaction of many motors. Early research along this line
was motivated by mRNA translation that is managed by ribosomes.
Ribosomes are bound to the mRNA strand with one subunit and step
forward codon by codon. The codon information is translated into
corresponding amino acids that are taken up from the cytoplasm and
assembled into proteins. To increase the protein synthesis many
ribosomes can be bound to the same mRNA strand simultaneously. This
fact might induce collective properties as was first realized by
MacDonald \cite{macdonald} who set up a theoretical model for the
translation of highly expressed mRNA. The importance of effects caused
by the concerted action of many motors can be deduced from a very
simple example that has yet drastic consequences: the slow down of
ribosomes due to steric hindrance caused by another ribosome in front
-- comparable to an intracellular traffic jam that might significantly
slow down protein synthesis.

A theoretical approach to collective phenomena in intracellular
traffic will try to simplify the processes of molecular motion
down to a single step rate rather than focus on the chemical or 
mechanical details
on the molecular level of motor steps. Then it becomes
possible to model and analyze the behavior of several motors with the
tools of many-body and statistical physics. We will start this review
with a short introduction on this single step model in Sec.
\ref{s_model} before we introduce the total asymmetric exclusion
process (TASEP) as a theoretical model for intracellular transport.
Sec. \ref{s_phase} describes the stationary states and density
distributions and their phase diagram as a function of boundary
conditions. After a review on several recent extensions in Sec.
\ref{s_extensions}, we will focus on the competition of TASEP and bulk
dynamics in Sec. \ref{s_pff}. Before concluding, Sec.
\ref{s_outlook} contains further recent developments.

\section{Model and Methods} \label{s_model}

In the quest for a theoretical model for the motion of molecular
motors the first and simplest choice may be the use of a Poisson
process. The ``Poisson stepper'' is assumed to be an extensionless
object advancing stochastically in discrete steps along a one-dimensional periodic
lattice. 
The process is uni-directional as the position of the stepper
can be described by $ x(t)=a \ n(t)$ with the discrete step size $a$ and
the random variable $n(t)$ being a sequence of growing integers. Step
events occur stochastically with a rate $r$ constant in both space and
time. Consequently, the average time between two steps is then given
by the dwell time $\tau = 1/r$ and the probability to find the
``Poisson stepper'' at a position
$n$ after time $t$ by the Poisson distribution \cite{vankampen}.

After we have defined a model for the translocation of a single motor,
we proceed with our original task which aims at the understanding of
collective properties of many motors. Of course, more elaborate
models have been established \cite{ps04,jp97} that account for several rate
limiting steps -- examples are the ATP supply or the availability of
amino acids for ribosomal mRNA translation. However, the very basic
``Poisson stepper'' is chosen for reasons of simplicity and in order to
prevent unnecessary molecular details from masking collective effects.
Still, the validity and limitations of this simplification have to be
kept in mind.

\begin{figure}
\centering
\includegraphics[width=\textwidth]{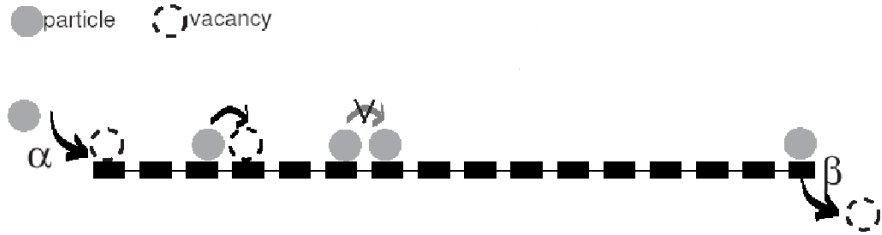}
\caption{Schematic model of TASEP (particles are injected with rate $\alpha$, move exclusively to
the right being subject to hard-core exclusion, and are removed with rate $\beta$)}
\label{tasepmodel}
\end{figure}

Being supplied with the dynamics of a single motor, a stage for the
concerted action of many can now be set up.  This was first done in a
pioneering work by MacDonald \cite{macdonald} and is now widely know
as the total asymmetric exclusion process (TASEP). It consists of an
one-dimensional lattice (Fig. \ref{tasepmodel}) with N sites labeled
by $i={1,\cdots,N}$ and with a spacing of $a=L/N$, where $L$ is the total
length of the lattice. For convenience, $L$ is often set to $1$ and
the lattice spacing then referred to as $\varepsilon=1/N$.

Particles have an extension of the size of the lattice spacing and are
subjected to hard core exclusion due to steric hindrance. Therefore
the occupation number $n_i$ of site $i$ can only take the values 0 or
1.  Particles on the lattice attempt jumps to their right neighboring
site with a rate $r$, which will be set to unity in the following.
Hereby, a reference time scale is set. The effective frequency of
jumps can be much smaller than $r$ when attempted jumps are rejected
due to an already occupied target site. The attempted jump rate to the
left is zero, since we deal with a total asymmetric exclusion process,
in contrast to the asymmetric exclusion process or the symmetric
exclusion process, where the jump rate to the left is non-zero or even
equal to the jump rate to the right.

Unless one uses periodic boundary conditions, specific dynamic rules
have to be defined at the boundaries, which play a crucial role in the
solution of the process. Among different other conditions (reflective,
open with a blockage) the most common type are open boundaries, which
we will use as well: at the left boundary ($i=1$) particles attempt to
attach with a rate $\alpha$, while they detach at the right boundary
($i=N$) with rate $\beta$. This is equivalent to two additional sites
$i=0$ and $i=N+1$ at the boundaries, which are connected to the system
by the bulk dynamics described above, and are constantly set to the
density $\alpha$ and $1-\beta$ respectively.

In spite of its simplicity, TASEP shows a wide range of interesting
properties. Since it was propelled into the scope of statistical
physicists,  
it has become a paradigm for non-equilibrium physics. In contrast to
equilibrium systems it lacks detailed balance but evolves into a
steady state where a non-vanishing current is maintained between
boundaries. Upon varying these boundary conditions, TASEP was found to
exhibit phase transitions which -- following general theorems
\cite{wagner} -- are not even allowed for one-dimensional equilibrium
systems in the absence of long-range interactions. However, the
analysis of non-equilibrium systems is considerably complicated by the
lack of universal concepts like the Boltzmann-Gibbs ensemble theory.
Feasible methods exist nevertheless and will be explained in the next
section.

\section{Density and Current in Stationary States} \label{s_phase}

In analyzing exclusion processes research can focus on a multitude of
different properties. The probably most obvious to address is the
density and current distribution in the stationary state. This is 
motivated by two reasons. On the one
hand, one intuitively attributes a strong importance to density
information with respect to the biological background as e.g. the
ribosome density is connected to the rate of protein synthesis. On the
other hand, promising experimental techniques can measure motor
densities and may allow for validation of theoretical models. Of
course, quite extensive research has also been attracted to a
multitude of different properties like correlation functions
\cite{de93,schuetz93}, relaxation properties \cite{ds00} or
super-diffusive spreading of fluctuations \cite{bks85} which will not
be the topic of this review. We will focus on analytical methods
(supported by numerical simulations) that are designed to investigate
spatial density distributions in the stationary state of the system.
To this end we will introduce some basic tools that have proven useful in
the exploration of TASEP properties. These are based on mean-field
approximations and reproduce many results that can also be derived exactly.
We are well aware that this approach neglects correlations as included
in the exact solutions that have been achieved for the TASEP density
profile by applying either recursion relations \cite{ddm92} or a
quantum Hamilton formalism with Bethe ansatz \cite{schuetz01}.

\subsection{Quantum Mechanics and Statistic Properties}

As an introduction we will outline some general statistical properties.
At any given moment, the system can be found in a certain
configuration $\mu$ made up of the occupation numbers at each lattice
site. The next occurring stochastic event (i.e. the jump of one
particle to a neighboring site) will therefore change the system to
another configuration $\mu'$. The transition probability $p_{\mu \to
  \mu'}$ is independent of the way the system had reached the initial
configuration. Since there is no memory of the system's history, but
any transition probability solely depends on the preceding state,
TASEP is a Markov process. In order to describe the system's
evolution, we can then use a master equation for the probability to
find the system in a certain state.
\begin{equation}
\frac{dP(\mu)}{dt}=\sum_{\mu' \neq \mu} 
\left[\omega_{\mu' \to \mu}P_{\mu'}(t)-\omega_{\mu \to \mu'}P_{\mu}(t)\right] \;,
\end{equation}
where the $\omega_{\mu \to \mu'}$ are the transition rates from one 
configuration $\mu$ to another $\mu'$. 

How can we now translate this general property into a description of
TASEP? To this end we will use a convenient notation, which applies
methods from the quantum mechanics toolbox in order to formulate the
master equation in terms of operators. It was introduced as ``quantum
Hamiltonian formalism'' and allows for exact solutions
\cite{schuetz01}.  We introduce operators $\hat n_i(t)$, which act as
occupation number operators, measuring the presence ($n_i=1$) or
absence ($n_i=0$) of a particle at site $i$. This results in the
Heisenberg equation (for an introduction see e.g. \cite{schuetz01})
\begin{equation}
\frac{d}{dt}\hat n_i(t)= \hat n_{i-1}(t)[1-\hat n_i(t)] - 
                         \hat n_i(t)[1-\hat n_{i+1}(t)] \;,
\label{heisenberg}
\end{equation}
where the first term on the right hand side constitutes the jump of a
particle from the left neighboring site to site $i$ (and thus a
particle gain) and the second term a jump from site $i$ to the adjacent
lattice site on the right (a particle loss). Note the intrinsic
exclusion constraint in both terms that prevents jump events if the
destination site is occupied, i.e. the expression in brackets equals
zero.  If one expresses these gains and losses in current, it becomes
convenient to use the current operator
\begin{equation}
\hat j_i(t)=\hat n_i(t)[1-\hat n_{i+1}(t)] \;.
\end{equation}
This allows to rewrite (\ref{heisenberg}) as a discretized form of a
continuity equation with the discrete divergence $\nabla \hat
j_i(t)=\hat j_i(t) - \hat j_{i-1}(t)$:
\begin{equation}
\frac{d}{dt}\hat n_i + \nabla \hat j_i(t)=0 \;.
\label{continuity}
\end{equation}
Similar equations for the boundaries are readily derived in the same
way.  Since we are interested in the average density on a certain
lattice site, we need to compute the time (or ensemble) average of the
operators. Equation (\ref{heisenberg}) gives an equation of motion for
the operator and allows to solve for the time evolution of $\langle
\hat n_i(t) \rangle$.  In executing the ensemble average of
(\ref{heisenberg}) two-point correlation functions like $\langle \hat
n_{i-1}(t)(1-\hat n_i(t)) \rangle$ appear on the right hand side. These
correlation functions again are connected to higher order correlations
via their equations of motion. The resulting infinite series of
correlation functions can be solved exactly for special cases only.
Generally, one is required to use mean-field approaches.

\subsection{Mean-Field Solution} \label{s_mf}
The rather blurry term ``mean-field theory'' is based on the concept of
using time or space averages (e.g. by neglecting temporal or spatial
fluctuations) and has found a wide range of applications in
statistical physics (see \cite{goldenfeld}). In this chapter, we
will explain its implementation for the TASEP and show a possible solution.
To point out the use of mean-field theory in TASEP, we look again at
the average of the operator $\hat n_i(t)$. We are interested in the
stationary state and therefore averaging signifies either a time or an
ensemble average, since the system is ergodic. Then the average
returns the density at the considered site $i$ as $ \varrho_i =
\langle \hat n_i \rangle $.  Performing the average over
(\ref{heisenberg}), leaves us with the difficulty of the infinite
series of correlation functions mentioned earlier. The mean field
approximation consists now in neglecting any correlations by setting
e.g. $\langle \hat n_i \hat n_j \rangle = \langle \hat n_i \rangle
\langle \hat n_j \rangle $ (see \cite{mahan}). In our case, we obtain
for the current
\begin{equation}
\langle \hat n_{i}(t)(1-\hat n_{i+1}(t)) \rangle = 
\langle \hat n_{i}(t) \rangle (1-\langle\hat n_{i+1}(t)\rangle) \;.
\end{equation}
This allows then, to rewrite (\ref{heisenberg}) in the stationary
state ($d \varrho_i(t)/dt=0$) as
\begin{equation}
0=\varrho_{i-1}(1-\varrho_i)-\varrho_i(1-\varrho_{i+1}) \;.
\label{e_steady}
\end{equation}
Obviously, (\ref{e_steady}) could easily be solved numerically, since
it forms a system of N difference equations.  However, it is possible
to reduce the set of equations to arrive at an explicit solution by
making two assumptions. First, note that the stationary state
condition $d \varrho_i(t)/dt=0$ implies a conservation of current
throughout the bulk, as can be seen from (\ref{e_steady}). Ergo, we
just have to solve one equation out of the set, to determine the
stationary, homogeneous current (and density).  For that purpose, we
use the continuum approximation, which turns the spatial lattice
variable quasi-continuous.  This is achieved for a large number $N$ of
lattice sites on a lattice of normalized length $L=1$.  The
distribution of sites then approaches a continuum as $\varepsilon=L/N
\ll 1$ and $x=i/N$ is rescaled to the interval $0 \leq x \leq 1$.
Thereby an expansion in powers of $\varepsilon$ is allowed.
\begin{equation}
\varrho(x \pm \varepsilon)=\varrho(x) \pm \varepsilon \partial_x
\varrho(x) 
+ \frac{1}{2}\varepsilon^2 \partial_x^2 \varrho(x) + O(\varepsilon^3)
\label{e_power}
\end{equation}
Using this expansion in (\ref{e_steady}) and the corresponding
equations for the boundaries, results in the following first-order
differential equation if we neglect all terms with higher orders in
$\varepsilon$:
\begin{equation}
(2 \varrho - 1) \partial_x \varrho  = 0 \;. 
\label{e_ode}
\end{equation}
The corresponding boundary conditions are $\varrho(0)=\alpha$ and
$\varrho(1)=1-\beta$. Because we have a first order differential equation
that needs to satisfy two boundary conditions, we are evidently
concerned with an over-determined boundary value problem. There are
three solutions to (\ref{e_ode}): $\varrho_\mathrm{bulk}(x) = 1/2$
does not satisfy either boundary condition (except for the special
case $\alpha=\beta=1/2$), while $\varrho(x)=C$ can satisfy either the
left or the right boundary condition, resulting in
$\varrho_\mathrm{\alpha}(x)=\alpha$ and
$\varrho_\mathrm{\beta}(x)=1-\beta$, respectively.

\subsection{Phase Diagram and Domain Wall Theory}

\begin{figure}
\centering
\includegraphics[height=4cm]{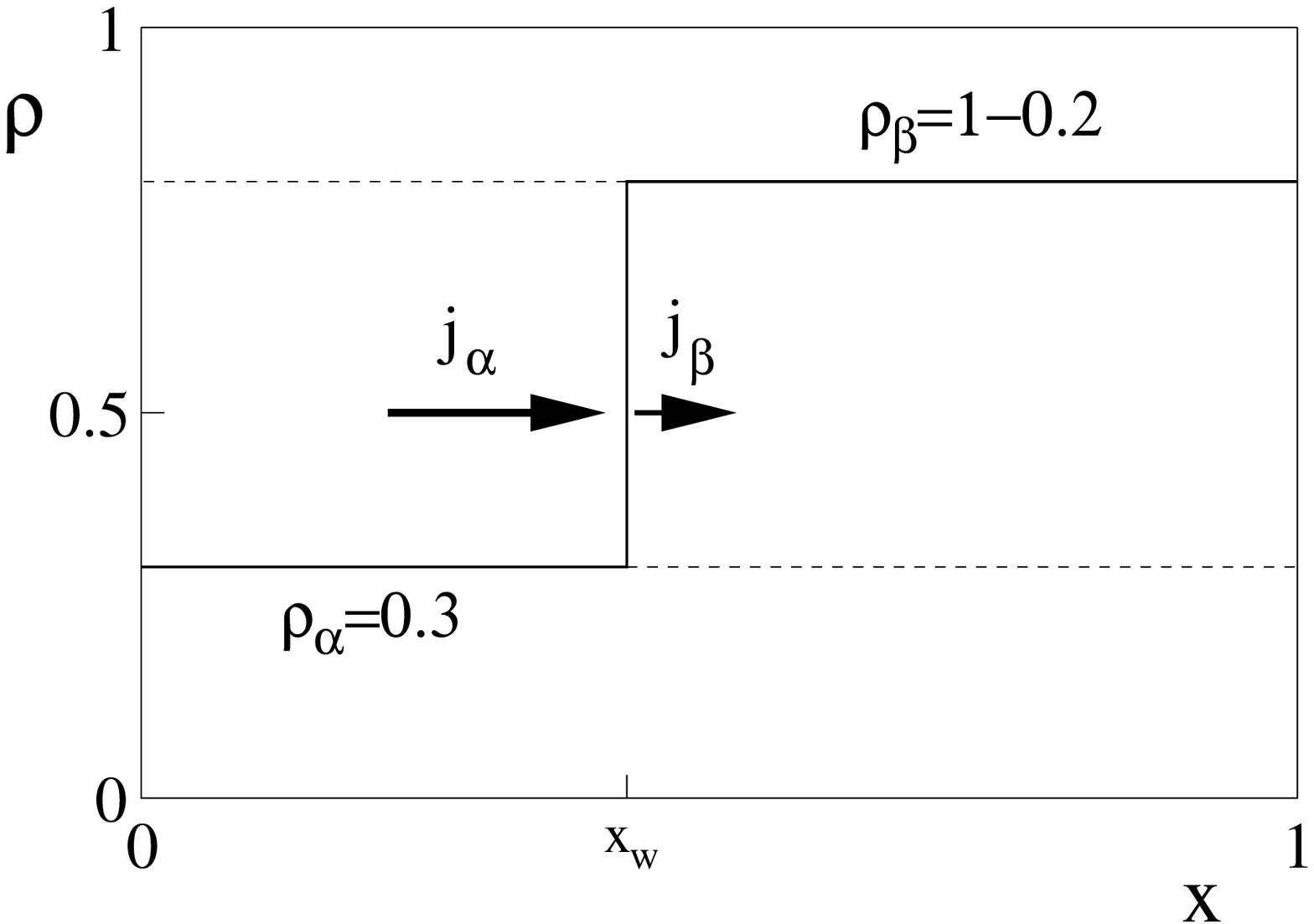}
\hspace{1 cm}%
\includegraphics[height=4cm]{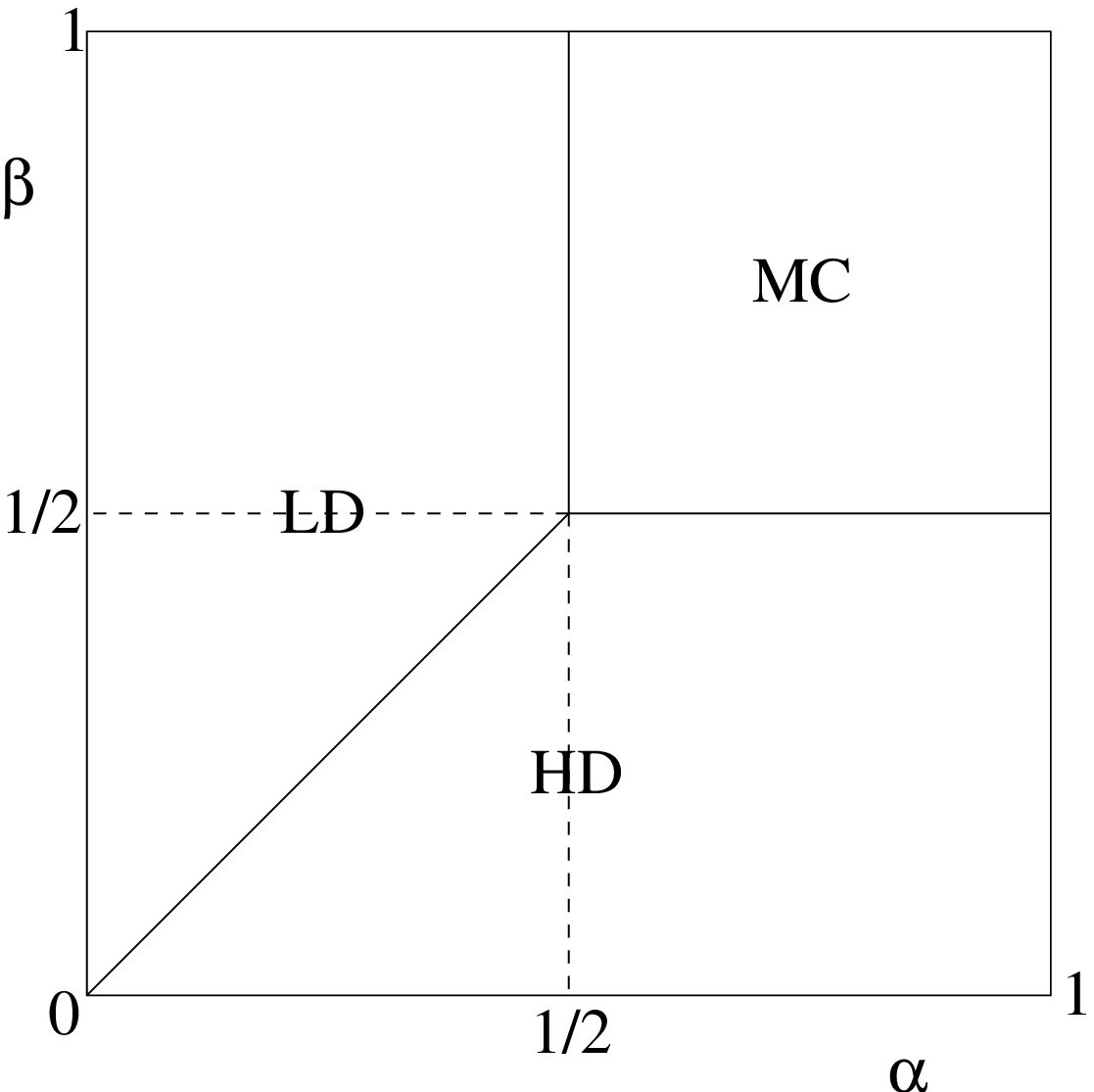}
\caption{({\bf left}) Schematic density distribution for 
  $\alpha=.3,\beta=.2$: in this situation the particle current
  $j_\mathrm{\alpha}$ exceeds the current $j_\mathrm{\beta}$, thus
  carrying more particles to the domain wall which is then shifted to
  the left. ({\bf right}) Phase diagram of TASEP in $\alpha,\beta$
  phase space shows a low density (LD), high density (HD) and maximal
  current (MC) phase}
\label{f_matching}
\end{figure}

To obtain a general solution satisfying the boundary conditions, both
solutions need to be matched.  Consider the case $\alpha,\beta < 1/2$.
To meet the boundary condition at both sides, the global density
function $\varrho(x)$ has to be $\varrho_\mathrm{\alpha}$
($\varrho_\mathrm{\beta}$) in an environment close to the left (right)
boundary. Since both $\varrho_\mathrm{\alpha}$ and
$\varrho_\mathrm{\beta}$ are uniform, the two solutions do not
intersect. At this point, we have to go beyond mean-field theory and
assume that at any given time both solutions are valid in
non-overlapping areas of the system. Where those areas border, they
are connected by a sharp domain wall (DW) at position $x_\mathrm{w}$
(see Fig.~\ref{f_matching})
\begin{equation} \label{e_dw}
\varrho(x)= 
\cases{ \varrho_\mathrm{\alpha} & for $0 \leq x \leq x_\mathrm{w} \;,$ \cr
                \varrho_\mathrm{\beta} & for $x_\mathrm{w} \leq x \leq 1 \;.$}
\end{equation}
From the dynamics of this domain wall \cite{kolomeisky-etal:98} we can deduce
important information about the system. The key point of domain wall
theory is the identification of particle currents as the cause of
domain wall motion. If for example the current
$j_\mathrm{\alpha}=\varrho_\mathrm{\alpha}(1-\varrho_\mathrm{\alpha})$
of the left density solution $\varrho_\mathrm{\alpha}$ is higher than
the current in the right part of the system (corresponding to
$\alpha>\beta$), particles are transported faster to the DW from the
left end then they can head on to the right. Thus, the domain wall is
shifted to the left. The system is filled up, until finally the whole
bulk density has taken the value of $\varrho_\mathrm{\beta}$ except
for a small boundary layer \footnote{The boundary layer is necessary
  in order to fulfill the both boundary conditions. Its extend is
  finite for small systems, but will vanish in the thermodynamic limit
  $N \to \infty$.}  at the system's left boundary. The opposite
happens, if $j_\mathrm{\beta}>j_\mathrm{\alpha}$, which is the case
for $\beta<\alpha$. Hence we can empirically state that the boundary
with the smaller rate acts as a bottleneck and imposes its density
distribution on the system.  To quantify this behavior, one can use a
traveling wave solution \cite{kolomeisky-etal:98} of the form
$\varrho(x-Vt)$ to obtain the domain wall velocity as
\begin{equation}
V = \frac{j_\mathrm{\beta}-j_\mathrm{\alpha}}
    {\varrho_\mathrm{\beta}-\varrho_\mathrm{\alpha}}=\beta - \alpha \;.
\end{equation}
To arrive at a phase diagram in $\alpha,\beta$-space we will analyze
the bulk density.  A positive DW velocity is obtained for
$\alpha<\beta$ and pushes the DW to the right side of the system
resulting in a bulk density smaller $1/2$ called low-density phase
(LD). On the contrary, $\beta<\alpha$ leads to a high-density phase
(HD), as illustrated in the phase diagram (Fig.~\ref{f_matching}). LD
and HD phase are connected by a first-order transition. Note, that
this phase diagram is equally obtained by the
above mentioned exact methods.

The transition line $\alpha=\beta<1/2$ in the phase diagram requires
special treatment. The velocity of the domain wall yields zero here,
but Monte Carlo simulations show that the DW actually will make random
steps to either side.  This behavior is caused by the stochasticity of
the input and output events that are described as rate processes.
Hence, the DW makes random steps to either side with equal
probability, being nothing else than the famous random walk in a
domain with reflecting boundaries.  An average over a sufficiently
long time will therefore result in a homogeneous probability density
over space. The stationary density profile will just be a linear slope
connecting the two boundary conditions:
\begin{equation}
\varrho(x)=\alpha+(1-\beta-\alpha)x \;.
\end{equation}
Note that mean-field theory neglects fluctuations and hence fails to
predict this density distribution, but gives the result of a
stationary domain wall. However, coming back to the picture of the
domain wall as a random walker, the linear density distribution can
readily be made plausible. Interpreting the random walk as free
diffusion of the domain wall and keeping in mind that the density
constraints $\varrho(0)=\alpha$ and $\varrho(1)=1-\beta$ hold at all
times, the linear density profile is easily derived as solution of
the one dimensional diffusion equation with boundary conditions.

Finally, consider an increase of e.g. $\alpha$ to values that exceed
$1/2$, while $\beta>1/2$ (crossing the boundary from the upper left to
the upper right quadrant in the phase diagram).  In this situation
particles are removed sufficiently fast from the system and the supply
of particles on the left side acts as a bottleneck limiting density
and current. For $\alpha<1/2$ any increase in $\alpha$ results in an
increased current (think of a highway, where an additional car will
result in a higher overall traffic). But above a certain value
$\alpha_\mathrm{C}=1/2$ any increase in particle input will not
increase the current further \footnote{As
  $j=\varrho(1-\varrho)=\varrho-\varrho^2$ has a maximum at
  $\varrho=1/2$. This density-current relation is a convenient tool to
  characterize traffic processes and its plot is often referred to as
  fundamental diagram.} but will cause the current to
diminish again (as an additional car will further slow down traffic
during rush hour).  As a result the bulk will keep its current maximum
at $\varrho=1/2$ and a boundary layer will form at the left side of
the system to match the boundary condition. This is of course nothing
else than the bulk solution $\varrho_\mathrm{bulk}$ with two boundary
layers and was baptized maximal current phase (MC). It is reached via
second-order transitions for values $\alpha>1/2$ and $\beta>1/2$. A
more rigorous treatment of this behavior can be gained by computing
the collective velocity of the particles \cite{kolomeisky-etal:98}.

\section{Biologically Motivated Generalizations of TASEP} 
\label{s_extensions}

The adoption of lattice gas models for biological systems was followed
by a variety of efforts to fit TASEP to different realistic
environments. For that purpose some of the simplifying assumptions
TASEP is based on had to be questioned. For example, experimental
observations have shown that ribosomes typically cover an area on the
mRNA that exceeds the lattice spacing by a multiple. To account for
this situation the particles in TASEP have to extend over several
lattice sites. However it was found that this does not change the
phase diagram quantitatively \cite{szl03}. Another direction of
research went back to the original field of MacDonald to elucidate the
importance of initiation and prolongation of ribosomes \cite{heinrich}
or formation of mRNA loops to facilitate the back transport of
ribosomes from the termination site \cite{chou03}.

While particle interactions in TASEP is limited to hard-core
potentials, even a small increase in the interaction radius -- as it
could be caused by charged molecules -- leads to qualitative changes in
the phase diagram \cite{ps99}.  It was shown that lattice gases with
short-range repulsive interaction exhibit a density-current relation
with two local maxima in contrast to simple TASEP that leads to one
maximum. This behavior results in a qualitative change of the phase
diagram that is enriched by four more regions one being a
minimal-current phase.

Further work was dedicated to the scenario of the interaction of
different species of molecular motors that move in opposite
directions. This can either happen on the same filament when two
different particles are able to surpass each other with a jump rate
that differs from the jump rate of either particle to a free site
\cite{efgm95} or on two adjacent one-dimensional filaments
\cite{pp01}. In both cases, spontaneous symmetry breaking was
observed.

Intracellular transport along cytoskeletal filaments has also served
as a source of inspiration for driven lattice gas models. While in the TASEP
model motors can only bind and unbind on the left and the right
boundary respectively, cytoskeletal motors are known to detach from the track
to the cytoplasm \cite{howard} where they perform Brownian motion and 
subsequently reattach to the track. The interplay between diffusion  
in the cytoplasm and directed motion along the filament was studied \cite{lkn01} 
both in open and closed compartments, focussing on anomalous drift and diffusion 
behavior, and on maximal current and traffic jams as a function of the 
motor density.

In \cite{pff03} it has been realized that the on-off kinetics may not only 
give rise to quantitative changes in the transport efficiency but also to 
a novel class of driven lattice gas models. It was shown that the interplay 
between bulk on-off kinetics and driven transport results in a stationary 
phase exhibiting phase separation. This was achieved by an appropriate 
scaling of the on-off rates that ensures that particles travel a finite 
fraction on the lattice even in the limit of large systems. Then, particles 
spend enough time to ``feel'' the their mutual interaction and, eventually, 
produce collective effects. In the following section, we will review the 
results of these studies \cite{pff03,pff04}.

\section{Phase Coexistence} \label{s_pff}

The essential features of cytoskeletal transport are the possibility of 
bulk attachment and detachment and a finite residence time on the lattice. 
The latter can be understood as a effect of thermal fluctuations that may overcome the
binding energy of the motors that is only of the order of several
$k_\mathrm{B} T$. Hence, attachment and detachment is a stochastic
process whose dynamic rules have to be defined.  Parmeggiani {\it et
  al.} \cite{pff03} chose to use Langmuir kinetics (LK) known
as adsorption-desorption kinetics of particles on a one- or
two-dimensional lattice coupled to a bulk reservoir \cite{vilfan}.
Particles can adsorb at empty sites and desorb from occupied sites and
microscopic reversibility demands that the kinetic rates obey detailed
balance leading to an evolution towards an equilibrium steady state
describable by standard concepts of equilibrium statistical mechanics.
In this sense the choice of LK is especially tempting as we are now
faced with the competition of two representatives of both equilibrium
and non-equilibrium systems.  The system -- in the following referred
to as TASEP/LK -- is defined as follows: the well-known TASEP is
extended with the possibility of particles to attach to the filament
with rate $\omega_\mathrm{A}$ and to detach from an occupied lattice
site to the reservoir with rate $\omega_\mathrm{D}$.  According to the
type of ensemble (canonical, grand canonical) the reservoir is either
finite or infinite.  Here, the reservoir is assumed to be infinite and
homogeneous throughout space and time. The density on a lattice
reached in the equilibrium state of LK is only dependent on the ratio
$K=\omega_\mathrm{D}/\omega_\mathrm{A}$ and is completely uncorrelated
in both space and time for neglection of any particle interaction
except hard-core. This is justified by the assumption that the
diffusion in the cytoplasm is fast enough to flatten any deviations of
the homogeneous reservoir density.  The resulting density profile on
the lattice is homogeneous and given as Langmuir isotherm
$\varrho_\mathrm{L}=K/(K+1)$.

\begin{figure}
\centering
\includegraphics[width=\textwidth]{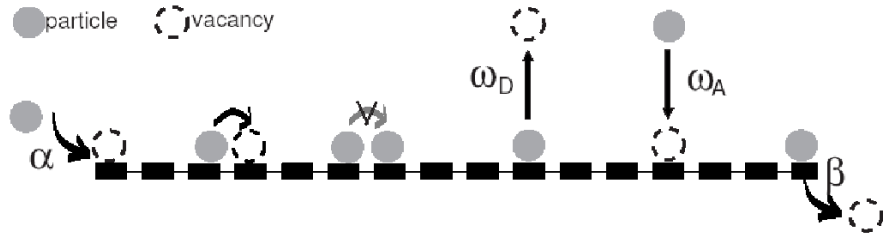}
\caption{Schematic model of TASEP/LK: 
  the TASEP is extended by possible particle attachment and detachment
  in the bulk with rate $\omega_\mathrm{A}, \omega_\mathrm{D}$}
\label{f_pff}
\end{figure}
If we now consider the combination of TASEP and LK into the model
displayed in Fig.~\ref{f_pff}, attention has to be paid to the
different statistical nature of both processes. TASEP evolves into a
non-equilibrium state carrying a finite current. Since particles are
conserved in the bulk, the system is very sensitive to the boundary
conditions, whereas LK as a equilibrium process is expected to be
robust to any boundary effects especially for large systems. Combining
both processes would thus lead to a trivial domination of LK as the
bulk rates $\omega_\mathrm{A}$ and $\omega_\mathrm{D}$ that apply to a
large number of bulk sites become predominant over the rates $\alpha$
and $\beta$ that only act on the two boundary sites.  To observe any
interesting behavior (i.e. real interplay) between the two dynamics,
one needs competition. A prerequisite for the two processes to compete
are comparable jump rates.  To ensure that the rates are of the same
order independently of the system size, an appropriate scaling is
needed. To this end a $N$-independent global detachment rate
$\Omega_\mathrm{D}$ is introduced, while the local rate per site
scales as
\begin{equation}
\omega_\mathrm{D}=\frac{\Omega_\mathrm{D}}{N} \;.
\label{e_scaling}
\end{equation}
For the attachment one proceeds similarly.  What does this scaling
signify physically? For an explanation it is instructive, to have a
look at the time scales involved: a particle on the lattice will
perform a certain move on an average time scale, which is the inverse
of that moves rate. Therefore, a particle spends an average time $\tau
\approx 1/\omega_\mathrm{D}$ on the lattice before it detaches.
Bearing in mind that the TASEP jump rate is set to unity, a particles
will jump to its adjacent site after a typical time of one unit time
step. Therefore the particle will travel a number $N_\mathrm{T} =
1/\omega_\mathrm{D}$ of sites before leaving the lattice. Compared to
the lattice length this corresponds to a fraction of
$n_\mathrm{T}=N_\mathrm{T}/N=1/(N \omega_\mathrm{D})$. In order to
keep this fraction finite in the thermodynamic limit,
$\omega_\mathrm{D}$ needs to scale as defined in (\ref{e_scaling}).
Only if the fraction $n_\mathrm{T}$ is finite, a given particle can
experience interaction with other particles and give rise to
collective phenomena \cite{pff04}. 

\subsection{Mean Field Solution of TASEP/LK}

To obtain density and current distributions of the TASEP/LK, we use
again a mean-field approach and proceed along the lines of Sec.
\ref{s_mf}.

First of all, we need to account for the Langmuir kinetics. This is
done by adding the following terms to the Heisenberg equation
(\ref{heisenberg}) of the simple TASEP to obtain
\begin{equation}
\frac{d}{dt}\hat n_i(t) = 
\hat n_{i-1}(t)(1-\hat n_{i}(t))-\hat n_{i}(t)(1-\hat n_{i+1}(t))
-\omega_\mathrm{D} \hat n_i(t)+\omega_\mathrm{A} (1-\hat n_i(t)) \;.
\end{equation}
where the first added term captures the detachment events and the
later the attachment.  Neglecting correlations as done before gives
for the stationary state
\begin{equation}
0 = \varrho_{i-1}(1-\varrho_i)-\varrho_i(1-\varrho_{i+1})-
    \omega_\mathrm{D} \varrho_i + \omega_\mathrm{A} (1-\varrho_i) \;.
\label{e_steady_pff}
\end{equation}
The boundary conditions are not altered by LK. Using again the power
series expansion (\ref{e_power}) and keeping in mind the scaling of
the on and off rates $\omega$ as in (\ref{e_scaling}), we obtain the
following ODE:
\begin{equation}
\frac{\varepsilon}{2}\partial_x^2 \varrho + \partial_x \varrho (2
\varrho - 1) - 
\omega_\mathrm{D} \varrho + \omega_\mathrm{A} (1-\varrho)=0 \;.
\label{e_ode_pff}
\end{equation}
The ratio $K=\omega_\mathrm{A}/\omega_\mathrm{D}$ between the
attachment and detachment rates will prove an important parameter in
the analysis of this differential equation. Since the case $K \neq 1$
is considerably complicated in its mathematical analysis we refer the
reader to reference \cite{pff04} and restrain our discussion to the
case $K=1$.  In the thermodynamic limit ($\varepsilon \to 0$) and with
$\omega_\mathrm{D}=\omega_\mathrm{A}=\Omega$ the ODE(\ref{e_ode_pff})
simplifies to first order:
\begin{equation}
(\partial_x \varrho - \Omega)(2 \varrho - 1)=0 \;.
\label{e_ode_pff3}
\end{equation}
Obviously, there are two solutions to this general ODE problem. The
homogeneous density $\varrho_\mathrm{L}=1/2$ given by the Langmuir
isotherm and the linear slope $\varrho(x)=\Omega x + C$. The value of
C is determined by the boundary conditions and leads to the two
solutions $\varrho_\mathrm{\alpha}(x)=\alpha + \Omega x$ and
$\varrho_\mathrm{\beta}(x)=1-\beta-\Omega + \Omega x$.  The complete
density profile $\varrho(x)$ is the combination of one or several of
the three densities above.  Depending on how they are matched, we
distinguish several phases as explained in the following.

\subsection{Phase Diagram and Density Distributions}

The only area in the phase diagram that does not change compared to
simple TASEP is the upper right quadrant. This is not surprising since
the maximal current phase is a bulk controlled regime anyway.
Therefore the additional bulk dynamics with the Langmuir isotherm at
$\varrho_\mathrm{L}=1/2$ do not result in any changes in the density
distribution. In this case, non-equilibrium and equilibrium dynamics
do not compete but cooperate. 

\begin{figure}
\centering
\includegraphics[height=7cm]{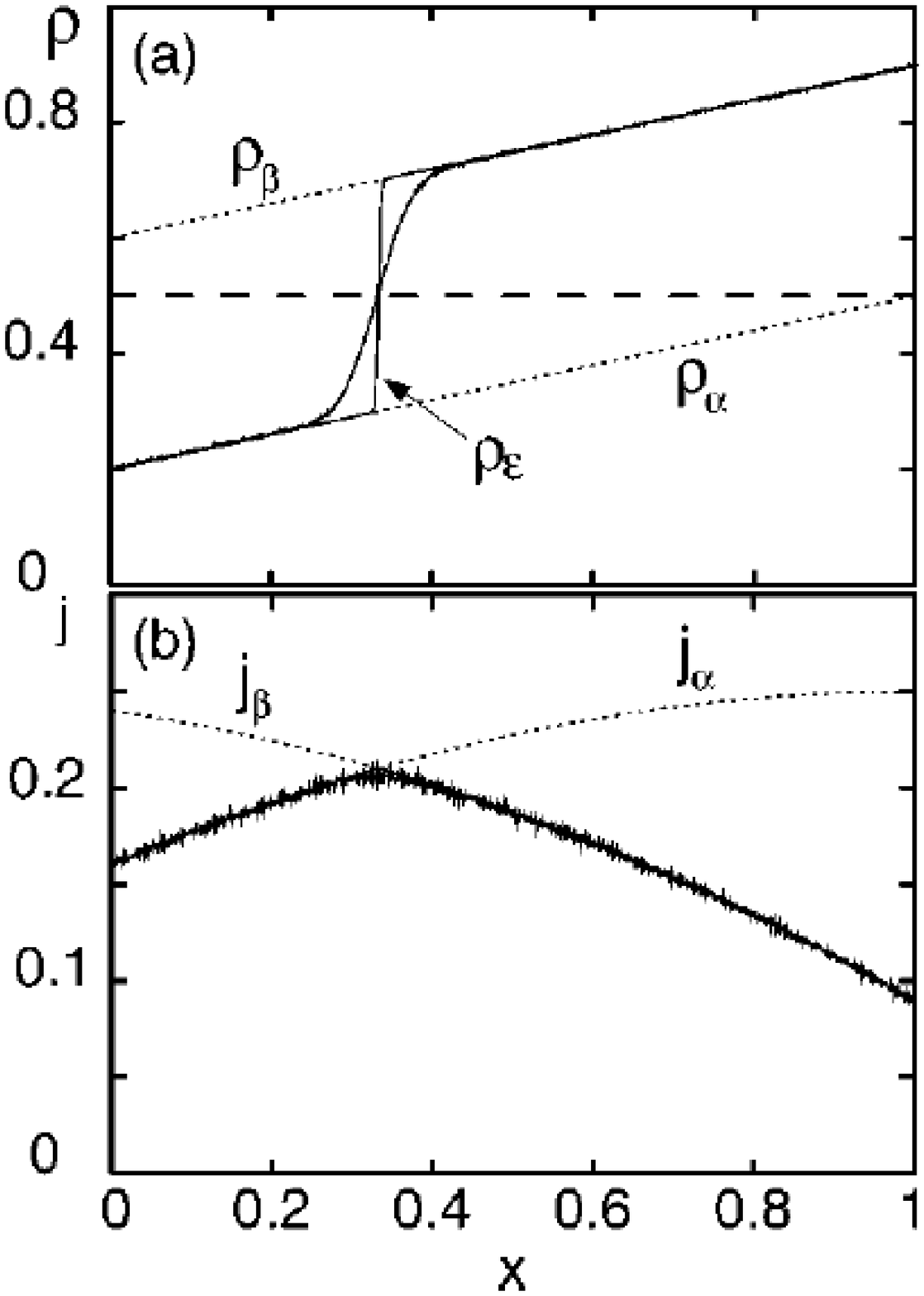}
\hspace{1 cm}%
\includegraphics[height=7cm]{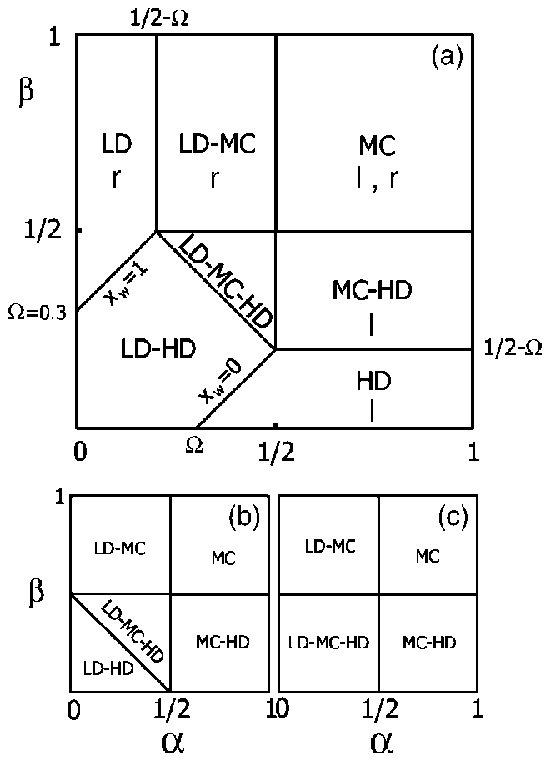}
\caption{({\bf left}) The DW connects the two densities $\rho_\alpha$ 
  and $\rho_\beta$ ({\it both dashed}) and is localized at the point
  where the correspondent currents $j_\alpha$ and $j_\beta$ match.
  Note the finite extend of the DW (localization length) that is only
  produced for Monte-Carlo simulations ({\it solid wiggly line}) and
  is not captured by mean-field results ({\it solid line}).  ({\bf
    right}) Topological changes in the phase diagram of TASEP/LK for
  (a) $\Omega=0.3$, (b) $\Omega=0.5$, (c) $\Omega=1$, from
  \cite{pff04}}
\label{f_pff_phase}
\end{figure}

As in TASEP different solutions can be matched in various ways, the
simplest being the connection by a domain wall between the left
solution $\varrho_\mathrm{\alpha}$ and the right solution
$\varrho_\mathrm{\beta}$.  Depending on the current distribution, two
possibilities have to be distinguished. As both solutions are
non-homogeneous, the corresponding currents $j_\mathrm{\alpha}$ and
$j_\mathrm{\beta}$ will be strictly monotonic (Fig. \ref{f_pff_phase}
(left)).  If the currents equal each other inside the system at a
position $x_\mathrm{w}$, the DW is localized at this position, as a
displacement to either side would result in a current inequality that
drives the DW back to $x_\mathrm{w}$ (see Fig. \ref{f_pff_phase}
(left)). Ergo, the TASEP/LK exhibits multi-phase existence of low and
high density regions (LD-HD phase) in the stationary state on all
time-scales, opposed to TASEP where this behavior is only observed for
short observation times. Recently, this DW localization has been
observed experimentally \cite{nosc}.

If the matching of left and right current is not possible inside the
system, the known LD and HD phases are found. This is the case for one
boundary condition being considerably larger than the other and is
evidently depending quantitatively on the slope of the density
solutions. This slope is determined by the ratio of the TASEP step
rate and the bulk interchange rate $\Omega$. For large $\Omega$ any
density imposed by the boundary relaxes fast against the Langmuir
isotherm of $\varrho_\mathrm{L}=1/2$ resulting a steep slope of the
density profile.

This fact allows for the existence of two other phases with
multi-regime coexistence. We could imagine a scenario in which the
boundary imposed density solutions decay fast enough towards the
isotherm to enable a three-regime coexistence of low density, maximal
current and high density (LD-MC-HD phase). Furthermore, a combination
of a MC phase with a boundary layer on one side and a LD or HD region
of finite extend at the other boundary can be imagined. Not all these
phases will be realized for every value of $\Omega$. Instead, the
phase topology of two-dimensional cuts through the
$\alpha,\beta,\Omega$-phase space changes. An example is shown in Fig.
\ref{f_pff_phase} (right).

\subsection{Domain Wall Theory}
\label{s_dw}

After we have derived an analytical solution for the density profile
based on the mean-field differential equation (\ref{e_ode_pff}), we
now have a closer look on the domain wall and its stochastic
properties.

As mentioned before, the density profile exhibits a discontinuity at
$x_\mathrm{w}$ that is actually a finite size continuous transition
between the high and low density for systems of finite size.  Only
upon increasing the system size $N \to \infty$ a sharp transition
between the left and the right solution occurs. However, this is only
due to the fact that the lattice spacing decreases to $\varepsilon \to
0^+$. So compared to the lattice length $L$, usually normalized to
$1$, the domain wall is discontinuous, while on the length scale of
lattice sites it will still have a finite extend. This extend is an
intrinsic statistical feature and is usually referred to as
localization length.

The domain wall in its random walk behavior can be either subjected to
equal rates (unbiased) as in TASEP or the rates of movement to the
left/right can be different (biased). In general, the rates will not
only be different, but also depend on the space variable.
To begin with, we will show a way how to derive these rates by taking
into account fluctuations of particle number
\cite{kolomeisky-etal:98,santen-appert:02}. Consider a situation where
all events that can change the particle number ($\alpha,\beta,\Omega$)
have a typical time scale that is considerably larger than that of
jump processes on the lattice. In this case, the time between any
entry and exit events is so long, that the system has enough time to
``rearrange'' (to reach a temporary steady state) in between . Then it
is possible to identify jump rates $\omega_\mathrm{l}(x)$
($\omega_\mathrm{r}(x)$) for DW movement to the left (right) with the
overall rate for entry and exit of any particle at any site.
Specifically, if a particle enters the system, the DW is shifted to
the left by a distance of $\approx
\varepsilon/[\varrho_\mathrm{\beta}(x_\mathrm{w})-\varrho_\mathrm{\alpha}(x_\mathrm{w})]$.
Therefore the rate for the DW to move one lattice site to the left is
$\omega_\mathrm{l} =
\omega_\mathrm{entry}/[\varrho_\mathrm{\beta}(x_\mathrm{w})-\varrho_\mathrm{\alpha}(x_\mathrm{w})]$.

\begin{figure}
  \centering \includegraphics[height=3.5cm]{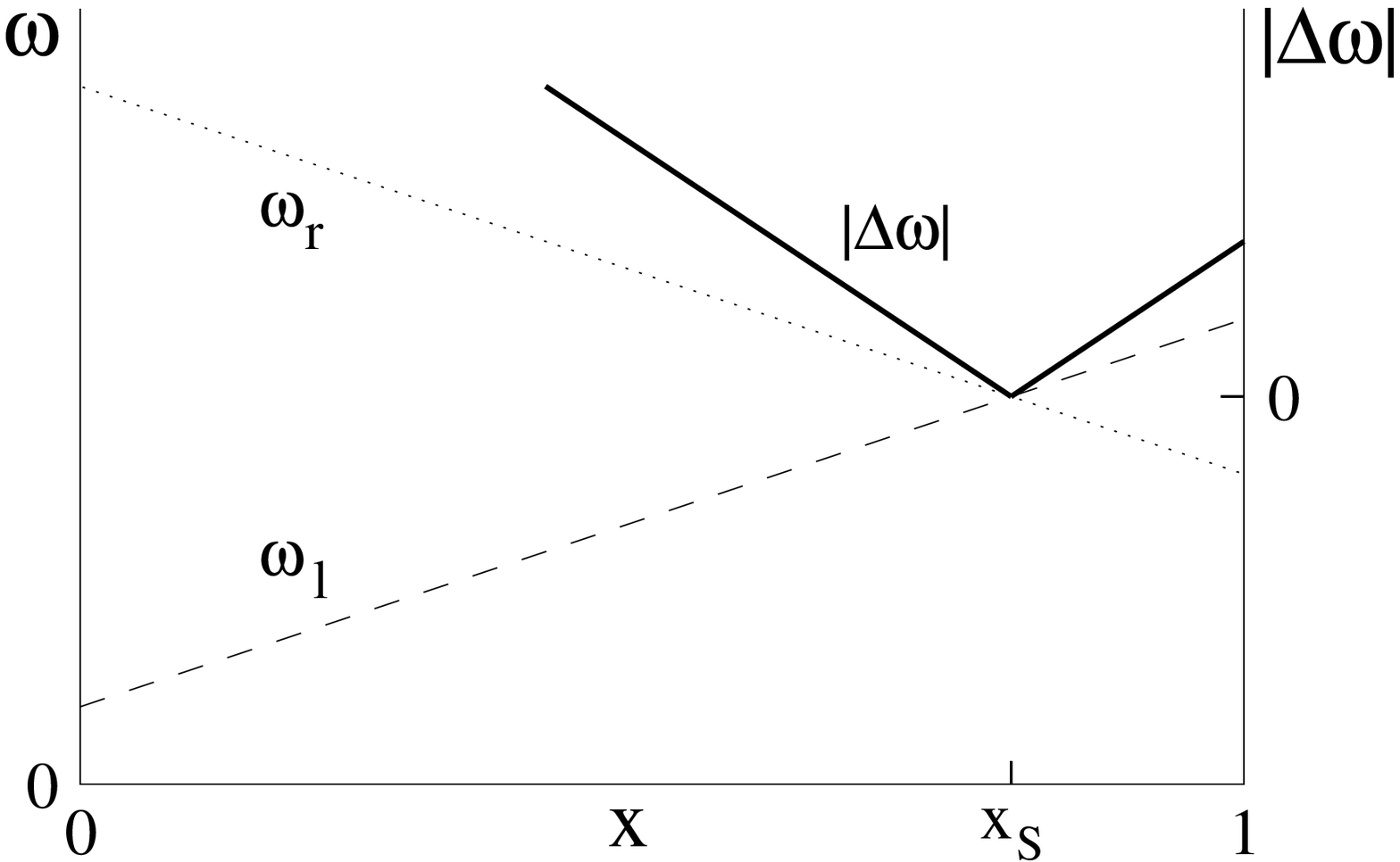}
\hspace{1 cm}%
\includegraphics[height=3.5cm]{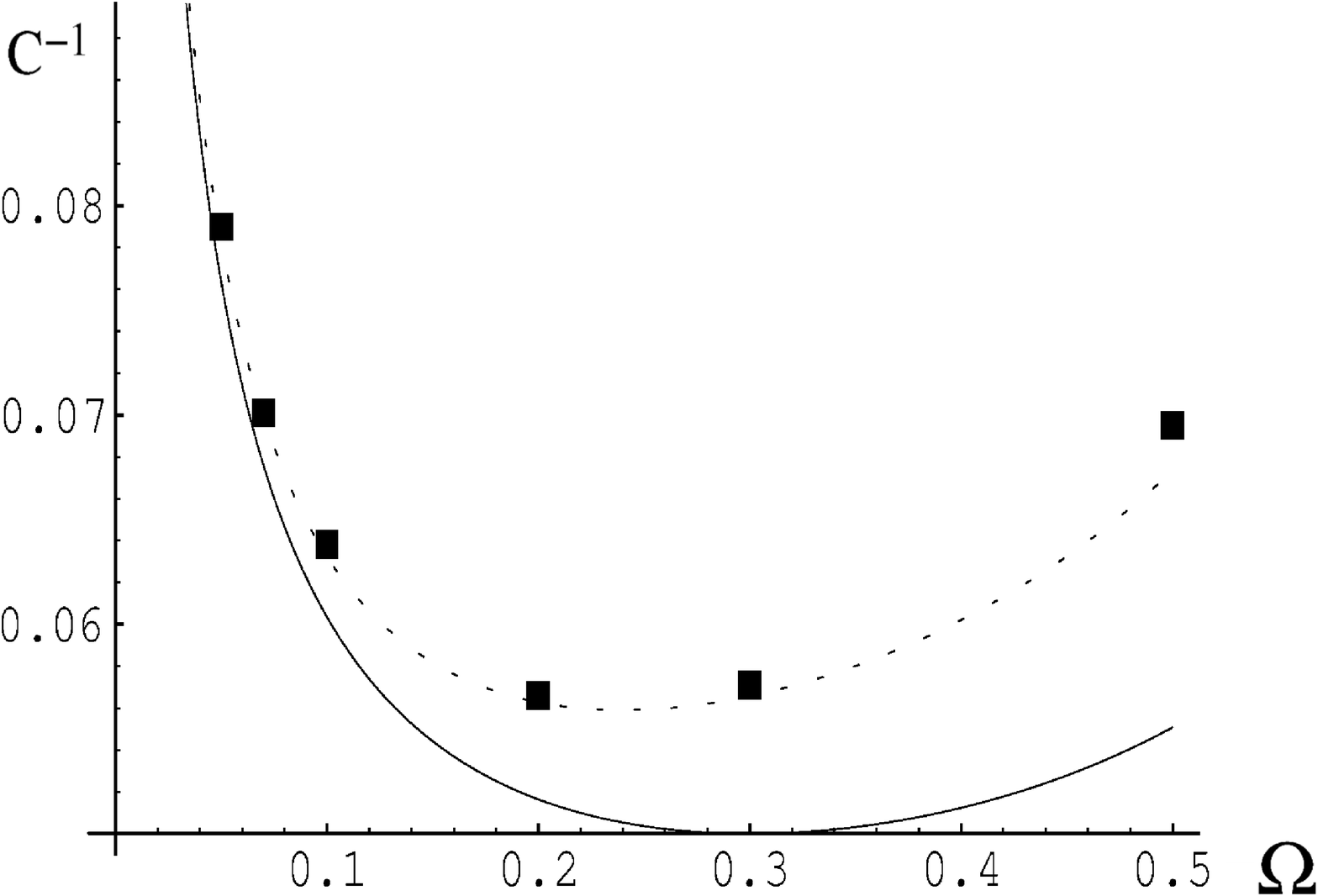}
\caption{({\bf left}) The DW will be localized at the point where the
  jump rate to the left $\omega_\mathrm{l}$ and right
  $\omega_\mathrm{r}$ intersect. This can be interpreted as an
  potential $|\Delta \omega|$. ({\bf right}) Variance of DW
  probability distribution over $\Omega$ for $\alpha=\beta=0.01$ and
  $\alpha=\beta=0.1$ in a system of $N=500$, predictions according to
  \cite{evans-juhasz-santen:03} ({\it solid}) and based on particle
  number fluctuations ({\it dashed}) compared to MC simulations ({\it
    dots})}
\label{f_potential}
\end{figure}

If the density distribution $\varrho(x)$ is known analytically, it is
possible to calculate $\omega_\mathrm{entry}$, the overall rate of
particle entrance, as the sum over all possible entrance events
\begin{equation} \label{e_partfluct}
\omega_\mathrm{entry}=\alpha(1-\alpha)+\int_0^1 dx \ \omega_\mathrm{A} (1-\varrho(x)) \;.
\end{equation}
The first term captures entrance events from the left boundary
reservoir and the integral accounts for the Langmuir kinetics. The
first multiplicand in both terms is the attempted rate of a jump,
whereas the difference in brackets states the probability of the
destination site to be vacant. Along the same lines, the exit rate is
computed.  As we know the analytical density distribution as
$\varrho(x)=\alpha+\Omega x + \Delta \Theta(x-x_\mathrm{w})$ with the
Heavyside function $\Theta$ and the DW height~\footnote{The term 
  $-1 \Omega$ accounts for the diminished height that is caused 
  by the Langmuir kinetics on the whole length $1$ of the system.} 
$\Delta=1-\alpha-\beta-\Omega$, we can execute the integrals
to obtain
\begin{equation}
\omega_\mathrm{entry}=\alpha(1-\alpha)+\Omega(\beta+\frac{\Omega}{2}) + x \Omega \Delta \;.
\end{equation}
Knowing these rates, one can complete the description of the domain
wall as a random walker by calculating the position dependent jump
rates to the left and right $\omega_\mathrm{l}(x)$ and
$\omega_\mathrm{r}(x)$. The two rates constitute an effective
potential displayed in Fig.~\ref{f_potential}.
The DW will always be driven to that point $x_\mathrm{S}$ in the
system where
$\omega_\mathrm{exit}(x_\mathrm{S})=\omega_\mathrm{entry}(x_\mathrm{S})$.
This position is the center of the DW probability density in a
stochastic picture and the stationary DW position in a mean-field
picture. It is in agreement with mean field results and yields
\begin{equation}
x_\mathrm{S}=\frac{\Omega-\alpha+\beta}{2\Omega} \;.
\end{equation}
The other quantity of interest which - in contrast to the DW position
- cannot be determined by mean-field calculations is the localization
length. In order to compute this quantity, fluctuations have to be
taken into account as we will show in the following.  If we use the
notation $p(x)$ for the probability that the domain wall is at
position $x$, then the condition for a stationary DW reads in the
continuum limit
\begin{equation}
\omega_\mathrm{r}(x)p(x)=\omega_\mathrm{l}(x+\varepsilon)p(x+\varepsilon) \;.
\label{e_stationary_dw}
\end{equation}
Introducing now $y(x)=\omega_\mathrm{l}(x)p(x)$ and approximate
$y'(x)=|y(x+\varepsilon)-y(x)|/\varepsilon$ we obtain:
\begin{equation}
y'(x)+ N y(x) (1-\frac{\omega_\mathrm{r}(x)}{\omega_\mathrm{l}(x)})=0 \;.
\end{equation}
The solution then is given by
\begin{equation}
p(x)=\frac{\tilde p(x)}{Z} = 
\frac{1}{Z \omega_\mathrm{l}(x)} exp[-N \int_{x_0}^x dx' (1-\frac{\omega_\mathrm{r}(x)}{\omega_\mathrm{l}(x)})] \;.
\label{e_p_dw}
\end{equation}
where $Z$ accounts for normalization. In general, $Z$ is not available
explicitly, but it has been shown \cite{evans-juhasz-santen:03} that
the unnormalized probability function can be approximated by a
Gaussian
\begin{equation}
\tilde p(x) \propto e^{-C(x-x_\mathrm{S})^2} \;,
\label{e_gaussian}
\end{equation}
where $C$ is given by the second order derivative of the exponent in
(\ref{e_p_dw}) as
\begin{equation}
C=\frac{1}{2} \frac{d^2}{dx^2}[N \int_{x_0}^x dx' \left(1-\frac{\omega_\mathrm{r}(x)}{\omega_\mathrm{l}(x)}\right)]=
\frac{N(\omega_\mathrm{l}-\omega_\mathrm{r})'(x_\mathrm{S})}{2 \omega_\mathrm{l}(x_\mathrm{S})} \;.
\label{e_C}
\end{equation}
Hence, the variance $\sigma=\sqrt{1/(2 C)}$ of the domain wall can be
easily be obtained provided that the jump rates $\omega_\mathrm{l}(x)$
and $\omega_\mathrm{r}(x)$ are available. Evans {\it et al.}
\cite{evans-juhasz-santen:03} have assumed those rates to be
$\omega_\mathrm{l,r}(x)=
j_\mathrm{\alpha,\beta}/(\varrho_\mathrm{\beta}-\varrho{\alpha})$.
Kouyos has shown \cite{roger} that using the rates (\ref{e_partfluct})
derived from the fluctuations of particle number, one arrives at more
accurate results compared to Monte Carlo simulations (see Fig.
\ref{f_potential}). In this case $C$ evaluates to
\begin{equation}
C=\frac{2 N \Omega \Delta}{\alpha(1-\alpha)+\beta(1-\beta)+\Omega} \;. 
\end{equation}
As the width of the DW is given by $\sigma=1/\sqrt{2C}$, the
localization of the DW scales with $N^{-1/2}$.

\section{Conclusions and Outlook} 
\label{s_outlook}

Much in the same way as MacDonald's pioneering paper \cite{macdonald} on
mRNA translation, recent work on kinesin motors walking along
microtubules has spurred progress in nonequilibrium transport
phenomena. The ubiquitous exchange of material between the cytoplasm
and the molecular track, which originally was thought to only lead to
quantitative modifications of the dynamics \cite{lkn01}, has recently
been identified \cite{pff03} as the source for qualitatively new phenomena such as
phase separation. This introduced a completely new class of
non-equilibrium transport models.

These lattice gas models are characterized by a scaling of the on-off 
rates with system size which enables competition of driven motion 
and equilibrium Langmuir kinetics. Hereby, a finite residence time on the 
lattice ensures cooperative effects to establish multi-phase coexistence, 
localized shocks and an enriched phase behavior compared to prior TASEP 
results.

There are now various routes along which one could proceed. The first
one is to add more realistic features of the molecular motors, such as
the fact that they are dimers \cite{pierobon} or
that there is more than one chemo-mechanical state \cite{nosc}. Such
investigations are crucial for a quantitative understanding of	
intracellular traffic in various ways.  One might ask how robust the
features of minimal models are with respect to the addition of more
molecular details. In the case of dimers the answer is far from
obvious since the non-equilibrium dynamics of dimer adsorption shows
rich dynamic behavior with anomalously slow relaxation towards the
equilibrium state \cite{vilfan}. How this combines with the
driven transport along the molecular track was recently analyzed
thoroughly \cite{pierobon}.  While correlation effects due to the extended
nature of dimers invalidate a simple mean-field picture it was found
that an extended mean-field scheme can be developed which
quantitatively describes the stationary phases. Surprisingly, the
topology of the phase diagram and the nature of the phases is similar
to the minimal model with monomers. The physical origin of this
robustness can be traced back to the form of the current-density
relation which exhibits only a single maximum.

The second line of research generalizing the minimal model \cite{pff03} asks
for the effect of interactions, more than one molecular traffic
lane, ``road blocks'' such as microtubule associated proteins and
various other kinds of ``disorder'',
bi-directional traffic, coupling of driven and diffusive transport 
and the like on the stationary density
profiles and the dynamics. In almost every instance it is found that
this leads to an even richer behavior with new phenomena emerging.

For TASEP it is known that isolated defects (slow sites) depending on
their strength may either give rise to local density perturbations for
low particle densities or yield macroscopic effects for densities
close to the carrying capacity \cite{tb98}.
The interplay between coupling to the motor reservoir in the
cytoplasm and the fluctuations in the capacity limit due to disorder
along the track gives rise to a number of interesting collective
effects \cite{kouyos}.

Coupling two lanes by allowing particle exchange at a constant rate
along the molecular track also results in novel phenomena. Similar to
equilibrium phase transitions described by field theories with two
coupled order parameters higher order critical points may emerge
\cite{reichenbach}.

Coupling diffusive and driven transport, the origin of new phenomena
is due to a competition of different processes of comparable time
scales. The qualitative failure of mean-field theory in some of these 
systems \cite{hinsch} comes as quite a surprise, since mean-field has proven to 
predict phase diagrams for large systems with an astonishing accuracy 
in the lattice gas models mentioned above.

\end{document}